\documentclass[journal]{IEEEtran}

\usepackage{amsmath}
\usepackage{amssymb}
\usepackage{setspace}
\usepackage{epsfig}
\usepackage{epstopdf}
\usepackage{rotating}
\usepackage{caption,subcaption}
\usepackage{color}
\usepackage{colortbl}
\usepackage{graphicx}
\usepackage{subfig}
\usepackage{ifthen}
\usepackage{alltt}
\usepackage{float}
\usepackage{fancyhdr}
\usepackage{verbatim}
\usepackage{moreverb}
\usepackage{cite}
\usepackage{multirow}
\usepackage{bigstrut}
\usepackage{subcaption}
\def\e{\begin{equation}}
\def\f{\end{equation}}
\def\_#1{{\bf #1}}
\def\o{\omega}

\def\E{\epsilon}
\def\M{\mu}

\def\.{\cdot}
\def\x{\times}
\def\Re{{\rm Re\mit}}
\def\Im{{\rm Im\mit}}
\def\l#1{\label{#1}}
\def\r#1{(\ref{#1})}
\def\=#1{\overline{\overline{#1}}}

\def\aeeb{\={\alpha}_{\rm ee}}
\def\aemb{\={\alpha}_{\rm em}}
\def\ameb{\={\alpha}_{\rm me}}
\def\ammb{\={\alpha}_{\rm mm}}
\def\aeeo{\alpha_{\rm ee}^{\rm co}}

\def\aemr{\alpha_{\rm em}^{\rm cr}}

\def\amer{\alpha_{\rm me}^{\rm cr}}
\def\ammo{\alpha_{\rm mm}^{\rm co}}

\def\aeeoh{\widehat{\alpha}_{\rm ee}^{\rm co}}
\def\aeerh{\widehat{\alpha}_{\rm ee}^{\rm cr}}
\def\aemoh{\widehat{\alpha}_{\rm em}^{\rm co}}
\def\aemrh{\widehat{\alpha}_{\rm em}^{\rm cr}}
\def\ameoh{\widehat{\alpha}_{\rm me}^{\rm co}}
\def\amerh{\widehat{\alpha}_{\rm me}^{\rm cr}}
\def\ammoh{\widehat{\alpha}_{\rm mm}^{\rm co}}
\def\ammrh{\widehat{\alpha}_{\rm mm}^{\rm cr}}
\def\aeebh{\={\widehat{\alpha}}_{\rm ee}}
\def\aembh{\={\widehat{\alpha}}_{\rm em}}
\def\amebh{\={\widehat{\alpha}}_{\rm me}}
\def\ammbh{\={\widehat{\alpha}}_{\rm mm}}

\def\me{\mathbf{E}_{\rm t}}
\def\mh{\mathbf{H}_{\rm t}}

\def\be{\beta_{\rm e}}
\def\bm{\beta_{\rm m}}
\def\n{\eta_0}
\def\=#1{\overline{\overline{#1}}}

\def\It{\={I}_{\rm t}}
\def\Jt{\={J}_{\rm t}}

\begin{document}

\title{Tailoring reflections from thin composite metamirrors}

\author{Y.~Ra'di,~\IEEEmembership{Student member,~IEEE}, V.~S.~Asadchy, and S.~A.~Tretyakov,~\IEEEmembership{Fellow,~IEEE}% <-this % stops a space
\thanks{The authors are with the Department of Radio Science and Engineering/SMARAD Center of Excellence, Aalto University, P.~O.~Box~13000, FI-00076 AALTO, Finland (e-mail: younes.radi@aalto.fi; viktar.asadchy@aalto.fi; sergei.tretyakov@aalto.fi).}
}

%dfsd

\markboth{Ra'di \MakeLowercase{\textit{et al.}}: Tailoring reflections from thin composite metamirrors}%
{Ra'di \MakeLowercase{\textit{et al.}}: Tailoring reflections from thin composite metamirrors}
\maketitle

%\today

\begin{abstract}
We propose an effective route to fully control the phase of plane waves reflected from electrically (optically) thin sheets. This becomes possible using engineered artificial full-reflection layers (metamirrors) as arrays of electrically small resonant bi-anisotropic particles. In this scenario, fully reflecting mirrors do not contain any continuous ground plane, but only arrays of small particles. Bi-anisotropic omega coupling is required to get asymmetric response in reflection phase for plane waves incident from the opposite sides of the composite mirror. It is shown that with this concept one can independently tailor the phase of electromagnetic  waves reflected from both sides of the mirror array.
\end{abstract}

%\tableofcontents

%%%%%%%%%%%%%%%%%%%%%%%%%%%%%%%%%%%%%%%%%%%%%%%%%%%%%%%%%%%%%%%%

\section{Introduction}
\l{sec:introduction}

\IEEEPARstart{W}{hile} the reflecting properties of mirrors and the focusing properties of lenses have been known since ancient times, general possibilities to tailor reflection and transmission of plane waves using thin metasufaces have been realized only recently. In what concerns the extended control over transmission, the transmitarray (e.g. \cite{transmit-array}) is the known technique based on the use of two parallel antenna arrays. This concept has been recently generalized as the meta-transmit-array in \cite{Alu}, where subwavelength (in the transverse plane) elements are used.  Another class of transmission-phase controlling layers is the phase-shifting surface \cite{McNamara}. Most of these structures contain several layers and are considerably thick in terms of the wavelength. But using various frequency-selective surfaces (e.g. \cite{FSS}) including inhomogeneous in the layer plane \cite{Capasso1,Capasso2}, transmission phase can be controlled also by electrically thin layers. Eliminating reflection while controlling transmission phase is possible using Huygens's metasurfaces \cite{Teemu,Grbic,Eleftheriades}. Some general limitations on realizing thin one-way transparent sheets have been published in \cite{Younes4}.

In this paper, we will focus on thin composite layers which fully reflect incident waves with a possibility to engineer their phase response. A known approach to control the reflection phase is the use of reflectarrays (e.g., \cite{Syrigos,Encinar}). Reflectarrays for microwave applications are usually realized as arrays of resonant metal patches over a metal ground plane. All the patches of the array usually have the same shape but slightly varying dimensions, so that the resonance frequency of the individual antenna elements varies to ensure the desired variations of the reflection phase over the array surface. Alternatively, varying reactive loading of the individual antenna elements can be used. Because of the use of resonant patches and considerable electrical distance between the array elements, homogenization of these layers is not possible, making realizations of conformal arrays on bent surfaces difficult. Other limitations are narrow frequency band and the presence of the metal ground plane which forbids transmission at all frequencies.

Other conventional structures which provide full reflection with engineered phase are artificial magnetic walls and  high-impedance surfaces (HIS). In contrast to electric walls (perfectly conducting surfaces), magnetic walls do not reverse the phase of reflected waves offering interesting possibilities for antenna and microwave applications.  Magnetic walls have a high surface impedance due to zero tangential magnetic field on their surface \cite{Collin,modeboo}. Corrugated conducting surface was one of the first designs introduced to exhibit a high surface impedance \cite{Elliot}. This is a transmission-line based structure utilizing many shorted quarter-wavelength transmission lines positioned close to each other, which transform the zero impedance at the end of the transmission lines to a very high reactive impedance at the open ends of the transmission-line sections. The main shortcoming of this structure is its large electrical thickness. In order to decrease the layer thickness, it is possible to fill the grooves with a high-permittivity dielectric or meander them. Later on, various corrugated surfaces have been utilized as artificial soft and hard surfaces \cite{Kildal}. Alternatively, to realize high reactive surface impedance, it is possible to use a parallel resonance with a capacitive frequency-selective surface, usually an array of metal patches (so called \emph{mushroom structure} \cite{Sievenpiper}).
This basic idea has been used by many researchers designing various structures realizing high surface impedance. Analytical models of these devices can be found in many papers \cite{dynamic,Clavijo,Luukkonen1,Luukkonen2}. The
mushroom layer HIS proposed by Sievenpiper \cite{Sievenpiper} has been used in many antenna applications: screens in mobile phones, absorbers, tunable antennas, for reducing thickness of steerable antennas, in reconfigurable antennas, e.g. \cite{Tretyakov,Gao,Costa1,Zhang,Padooru,Costa2,Deo,Huang}. In paper \cite{Pozar_dense_reflect_array} a high-impedance surface with a non-uniform distribution of the resonant frequency over the surface has been used as a reflectarray, with the advantage of a wider bandwidth as compared to the traditional designs as antenna arrays \cite{Syrigos,Encinar}. As for any linear and passive structure, there are fundamental causality limitations on the thickness and bandwidth of high-impedance surfaces \cite{Gustafsson}. The main feature which considerably limits the functionalities of the mushroom structure is the presence of an electric wall (ground plane). The structure interacts with a very large frequency spectrum, but the transmittance is always zero. Furthermore, the reflection coefficient seen from the side of the ground plane is always that of a conducting wall ($R=-1$ if losses are negligible), unless a multi-layer system is used.

%Interestingly, after Sievenpiper's design there has not been any new design to propose HIS and this design seems to be accepted as the only and optimum way to get HIS. The main focus in this field has been on making the HIS tunable \cite{Costa3} or wideband and multiband HIS by different patterning of the patches array \cite{Kern,Costa4,Chen}. If we look at the historical progress of magnetic walls, we see that all researchers have been using almost same idea and the physics behind all these designs are the same which has led them to the final design of HIS \cite{Sievenpiper}.

To realize a magnetic wall in one single sheet, one needs to have a sheet of magnetic current. There are no free magnetic charges in nature, and only using dense arrays of resonant magnetic moments it is possible to effectively create a sheet of magnetic current. Magnetic walls made of a dense array of single and double split-ring resonators have been proposed in \cite{SimovskiD} and \cite{Vegni}. This design does not include  any ground plane in its structure, but the possible functionalities are still limited because the reflection coefficient seen from the opposite sides of the sheet is always the same.

The main questions which we address in this paper are: What are the minimum necessary physical requirements for realizing a single-sheet full-reflection structure with different and controllable phases of reflected waves from its different sides? Are the conventional designs of high-impedance surfaces and reflectarrays the only and optimal ways to control the reflecting properties of thin layers? It is obvious that to be able to fully control reflection, we need to ensure that the thin layer is polarizable both electrically and magnetically. Thus, we assume that the thinnest possible realization is a planar array of small dipolar particles in which both electric and magnetic moments can be induced by the incident fields. From the earlier studies of dipolar arrays \cite{Teemu} it is known that in order to be able to realize different reflectivity properties for the opposite sides of the sheet, the induced electric and magnetic dipoles must be coupled as in a bi-anisotropic particle \cite{basic}. To the best of our knowledge, bi-anisotropy of conventional full-reflection layers with asymmetric response of their two sides has not been noticed in earlier studies. Furthermore, an interesting question is if it is possible to have a single layer of bi-anisotropic particles which offers different reflection phase distributions on its two faces (e.g., a uniform reflection phase when illuminated from one side and a non-uniform reflection phase distribution from the other side)? To address these questions, it is necessary to find out which physical phenomena govern the response of traditional full-reflection layers and understand what parameters of reflecting sheets are required to realize arbitrary reflection phase responses.

In this paper, we will analytically and numerically show that the most general asymmetric single-layer full-reflection structures can be realized as single arrays of passive lossless bi-anisotropic particles possessing bi-anisotropic omega coupling \cite{basic}. It will be shown that using the proposed metasurfaces one can tailor the reflected wave phase independently for illuminations coming from the opposite sides of the composite sheet. We will see that using the physical optics principle (also known as the  Kirchhoff approximation \cite{Kirchhoff}) the proposed method is applicable for layers with different non-uniform phase distributions on their two surfaces. This study will also lead us to discover other extreme properties of so called \emph{balanced bi-anisotropic particles} \cite{Younes1,Younes2,Younes3,Younes4}.
With the ability to fully control the phase of waves reflected from a planar composite sheet, it becomes possible to tailor reflected waves in the most general fashion. For example, one side of a thin planar sheet can act as a focusing reflector (emulating a parabolic mirror), while the other side can, for instance, deflect incident plane waves into a desired direction. This study is limited to the case of plane-wave illuminations at normal incidence.

\section{General formulas for reflected and transmitted fields}

We consider a generic metamirror as a composite planar layer formed
by periodically arranged electrically small unit cells or small
particles. These cells are polarizable both electrically and magnetically. The electromagnetic properties of the cells we model by
general linear relations between the induced electric ($\_p$) and magnetic ($\_m$) dipole moments
and the incident fields $\mathbf{E}_{\rm inc}$ and
$\mathbf{H}_{\rm inc}$ at the positions
of the particles as \e \left[ \begin{array}{c} \mathbf{p} \\
\mathbf{m}\end{array} \right] =\left[ \begin{array}{cc} \aeebh&
\aembh\\
\amebh& \ammbh \end{array} \right]\. \left[ \begin{array}{c}
\mathbf{E}_{\rm inc} \\ \mathbf{H}_{\rm inc}\end{array} \right] ,
\l{eq:a} \f where the effective polarizabilities (marked by hats)
include the effects of particle interactions in the array. Because
here we are interested in layers acting as full reflectors for
arbitrary polarizations of the incident field, the dyadic
coefficients in the above equations should be invariant with respect
to rotation around the unit vector $\_z_0$ normal to the layer plane: \e \={\widehat{\alpha}}_{ij}=\widehat{\alpha}_{ij}^{\rm
co}\=I_{\rm t}+\widehat{\alpha}_{ij}^{\rm cr}\=J_{\rm t},
\l{forms}\f where $\It=\={I}-\mathbf{z}_0\mathbf{z}_0$ is the
transverse unit dyadic and $\=J_{\rm t}=\mathbf{z}_0\x\={I}_{\rm t}$
is the vector-product operator. The amplitudes of plane waves generated
by the array are proportional to the surface-averaged electric and
magnetic surface current densities, which are related to the induced cell dipole
moments as $\_J_{\rm e}=j\omega \_p/S$, $\_J_{\rm m}=j\omega \_m/S$,
where $S$ is the area of each unit cell. Assuming the normal
incidence of the incident plane wave (the induced dipole moments are
the same in all unit cells), we can find the amplitudes of the
reflected and transmitted plane waves as \cite{Teemu} \e
\begin{array}{l}
\displaystyle
\mathbf{E}_{\rm r}=-\frac{j\omega}{2S}\left\{\left[\eta_0\aeeoh\pm \aemrh\pm \amerh-\frac{1}{\eta_0} \ammoh\right]\It\right.\vspace*{.2cm}\\\displaystyle
\hspace*{1.6cm}\left.+\left[\eta_0\aeerh\mp \aemoh\mp \ameoh-\frac{1}{\eta_0} \ammrh\right]\Jt\right\}\cdot\mathbf{E}_{\rm inc},
\vspace*{.3cm}\\\displaystyle
\displaystyle
\mathbf{E}_{\rm t}=\left\{\left[1-\frac{j\omega}{2S}\left(\eta_0\aeeoh\pm\aemrh\mp\amerh
+\frac{1}{\eta_0}\ammoh\right)\right]\It\right.\vspace*{.2cm}\\\displaystyle
\hspace*{.3cm}\left.
-\frac{j\omega}{2S} \left[\eta_0\aeerh\mp\aemoh
\pm\ameoh+\frac{1}{\eta_0} \ammrh\right] \Jt\right\}\cdot\mathbf{E}_{\rm inc},
\end{array}\label{eq:b}
\f where $\eta_0$ is the free-space wave impedance (for simplicity
and without loss of generality we assume that the surrounding
homogeneous, isotropic, and lossless medium is free space).
 Here and thereafter, to distinguish between illuminations of the sheet
from the two opposite sides, we will use double signs for these two
cases, where the top and bottom signs correspond to the incident
plane wave propagating in the $-\mathbf{z}_0$ and $\mathbf{z}_0$
directions, respectively. Ideal reflectors should not change the
wave polarization at reflection, so we demand that in (\ref{eq:b}) the coefficients at
the rotation operator $\=J_{\rm t}$ equal zero. In this work, we limit
the study to reciprocal layers, in which case the electric and
magnetic polarizabilities are symmetric dyadics ($\aeerh=0$,
$\ammrh=0$, e.g. \cite{basic}). Thus, the requirement of the absence
of any cross-polarized response does not allow any chirality in the
layer's topology, because both $\aemoh=0$ and
$\ameoh=0$.\footnote{If non-reciprocity is allowed, bi-anisotropic
chiral coupling can exist in the structure, but the effect of
chirality must be balanced with the effects of the Faraday rotation
due to non-zero antisymmetric parts of electric and magnetic
polarizabilities.} Under these restrictions, the amplitudes of reflected and
transmitted plane waves for normally incident plane waves read \e\begin{array}{l} \displaystyle \mathbf{E}_{\rm
r}=-\frac{j\o}{2S}\left(\n\aeeoh\pm \aemrh\pm \amerh-\frac{1}{\n}
\ammoh\right)\mathbf{E}_{\rm inc}, \vspace*{.3cm}\\\displaystyle
\mathbf{E}_{\rm t}=
\left[1-\frac{j\o}{2S}\left(\n\aeeoh\pm\aemrh\mp\amerh
+\frac{1}{\n}\ammoh\right)\right]\mathbf{E}_{\rm inc}.
\end{array}\l{eq:d}
\f

\section{Required polarizabilities for electric- and magnetic-wall response}
\label{PEC_PMC}

Before studying general metamirrors, let us first consider the simplest problem
of realizing electric or magnetic wall response for illumination
from both sides of the array.

\subsection{Electric-wall response}

If we require symmetric response on illuminations from both sides of
the array, then the field-coupling coefficients $\aemrh$ and
$\amerh$ in \r{eq:d} must be zero. As is obvious from \r{eq:d}, the array behaves as
an electric wall (reflection coefficient $R=-1$, transmission
coefficient $T=0$) if \e \aeeoh={2S\over j\o \n},\qquad
\ammoh=0.\l{eq:h}\f The effective electric polarizability $\aeeoh$
and the effective magnetic polarizability $\ammoh$, in terms of the
single-particle polarizabilities and the interaction constants,\footnote{Single-particle polarizabilities (notations without hats) connect the induced dipole moments with the local fields $\_E_{\rm loc}$ and $\_H_{\rm loc}$ at the position of the particle. The interaction constants define the values of the local fields as $\_E_{\rm loc}=\_E_{\rm inc}+\be\_p$ and $\_H_{\rm loc}=\_H_{\rm inc}+\bm\_m$.} can
be written as \e
 \aeeoh=\frac{\aeeo}{1-\aeeo\be}, \qquad \ammoh=\frac{\aeeo}{1-\ammo\bm},
\l{eq:i}\f where $\be$ is the electric-field interaction constant
and $\bm$ is its magnetic counterpart. Approximate analytical
expressions for the electric and magnetic interaction constants in
regular arrays of dipole particles can be found in \cite{modeboo}:
\e
\begin{array}{c}
\displaystyle
\be=\Re\left\{-\frac{jk_0}{4\epsilon_0 S}\left(1-\frac{1}{jk_0\rho}\right)e^{-jk_0\rho} \right\}
\vspace*{.2cm}\\\displaystyle\hspace*{.5cm}
+j\left(\frac{k_0^3}{6\pi\E_0}-\frac{k_0}{2\epsilon_0 S}\right)  \vspace*{.2cm}\\\displaystyle
\bm=\frac{{\beta}_{\rm e}}{\n^2}, \qquad
\rho=\frac{a}{1.438}
\end{array}\l{eq:j}
\f ($k_0$ is the free-space wave number and $a$ is the array period).
Here the expression for the imaginary part is exact, and that for
the real part is an approximation valid for $k_0a < 1$.

Considering the corresponding grid impedance $Z_{\rm g}$ \cite{modeboo} defined via
 \e \me=Z_{\rm g}\_J_{\rm e},\l{eq:k}\f where $\me$ is
the surface-averaged tangential electric field in the array plane
and $\_J_{\rm e}=j\o \_p/S$ is the surface-averaged induced electric
surface current density, we find that \e Z_{\rm g}={S\over j\o}
{1\over \aeeoh}-{\n\over 2}=0\l{eq:l}\f if \r{eq:h} is satisfied.
Using \r{eq:h} and \r{eq:i}, we get the conditions on the
polarizability of a single particle which is necessary to realize an
effective electric wall response: \e
\Re\left\{\frac{1}{\aeeo}-\be\right\}=0,\qquad
\Im\left(\frac{1}{\aeeo}\right)={k_0^3\over 6\pi\epsilon_0}.
\l{eq:m}\f The first condition means that the particles are at the
resonance and the second condition means that there is no absorption
in the particles (the only loss term is due to dipole scattering).
This is a known result
%\cite{}
telling that an array of resonant
lossless electric dipole particles behaves as an electric wall at
the particle resonance. The first equation also shows that the
resonance frequency is the frequency at which the grid comes to a
resonance, and this frequency is not the same as the resonance frequency of a
single particle in free space.

Let us compare the behavior of  a grid of resonant electric dipoles
with that of a thin homogeneous slab (the relative permittivity
$\epsilon_{\rm r}$ and the thickness $d$). For a thin homogeneous slab the equivalent sheet
impedance reads (eq. (2.152) \cite{modeboo})
\e Z_{\rm
g}={-j\eta_0\over (\epsilon_{\rm r}-1)k_0d},\l{eq:n}\f
and, obviously, the electric-wall
response ($Z_{\rm g}\rightarrow 0$) can be reached only if
$|\epsilon_{\rm r}|\rightarrow \infty$. If the imaginary part of
$\epsilon_{\rm r}$ tends to infinity, we have the case of a perfect
conductor, and if the real part of $\epsilon_{\rm r}$ tends to infinity,
we have the response analogous to the above case of an array of lossless
resonant electric dipoles. Indeed, in the case of the dipole array
the perfect electric conductor (PEC) response is reached when the
particles are at resonance and their polarizabilities in the grid
tend to infinity. After appropriate homogenization, this corresponds
to a thin layer of a dielectric with ${\rm
Re}(\epsilon_{\rm r})\rightarrow \infty$.

Note that a single sheet of homogeneous surface electric current
radiates symmetrically in the forward and back directions. In our
case of full reflection, the amplitude of the reflected plane wave
equals to that of the incident plane wave. The plane wave created by
this array into the forward direction also has the same amplitude as
the incident field, but it is out of phase with that, so that the
total field behind the grid is zero. Note that the discrete nature
of the grid is not essential for this interference effect
(cancellation of the incident field behind the grid). It is
essential in order to ensure the resonant response of the grid
impedance, which in a homogeneous layer is possible only with
$|\epsilon_r|\rightarrow \infty$.

\subsection{Magnetic-wall response}

Similarly to the above case of the electric wall, inspecting \r{eq:d}
we can find that the magnetic-wall response from both sides of the
array (the reflection coefficient $R=1$ and the transmission coefficient $T=0$)
is achieved if \e \aeeoh=0,\qquad \ammoh={2S\n\over j\o }.\l{eq:o}\f
The effective magnetic polarizability $\ammoh$ is related to the
polarizibility of the same particle in free space as in \r{eq:i}. We
can introduce the equivalent ``series admittance'' $Y_{\rm m}$ via \e\mh=Y_{\rm
m}\_J_{\rm m}\l{eq:q}.\f In this definition, $\mh$ is the
surface-averaged tangential magnetic field in the array plane and
$\_J_{\rm m}=j\o \_m/S$ is the surface-averaged magnetic
surface  current density. Similarly to the case of the electric-wall
response of resonant electric dipoles, we find \e Y_{\rm m}={S\over
j\o} {1\over \ammoh}-{1\over 2\n}=0.\l{eq:r}\f Thus, an array of
lossless resonant magnetic dipoles behaves as a magnetic wall from
both sides of the array at the resonant frequency of the array which
is determined, using \r{eq:i} and \r{eq:o}, by the following values of the particle polarizabilities: \e
\Re\left\{\frac{1}{\ammo}-\bm\right\}=0,\qquad
\Im\left(\frac{1}{\ammo}\right)={k_0^3\over 6\pi\mu_0}. \l{eq:s}\f

\subsection{Symmetric response with an arbitrary reflection phase}

Let us next consider if and how one can realize a sheet reflecting plane waves symmetrically from both sides, but with an arbitrary reflection phase. We use  \r{eq:d} with $\aemrh=0$ and
$\amerh=0$ to find the effective electric and magnetic polarizabilities of the unit cells which ensure that $\_E_{\rm t}=0$ and $\_E_{\rm r}=e^{j\phi}\_E_{\rm inc}$. Here $\phi$ is the desired phase shift of the reflected wave. The result reads
\e \eta_0\aeeoh={S\over j\omega}\left(1-e^{j\phi}\right),\qquad {\ammoh\over \eta_0}=
{S\over j\omega}\left(1+e^{j\phi}\right).\l{sym_a_hat}\f
The corresponding values of the required electric impedance and magnetic admittance for the surface-averaged current densities (see \r{eq:k} and \r{eq:q}) are given by
\e Z_{\rm g}={S\over j\o}
{1\over \aeeoh}-{\n\over 2}=j{\n\over 2}\cot(\phi/2)\f
\e Y_{\rm m}={S\over
j\o} {1\over \ammoh}-{1\over 2\n}={j\over 2\n}\tan(\phi/2)\f
The required polarizabilities of individual particles we find using \r{sym_a_hat} and \r{eq:h}:
\e \Re\left\{\frac{1}{\aeeo}-\be\right\}=-{\omega \eta_0\over 2S}\cot(\phi/2),\hspace{.1cm}
\Im\left(\frac{1}{\aeeo}\right)={k_0^3\over 6\pi\epsilon_0}.\l{aee_sym}
\f
\e \Re\left\{\frac{1}{\ammo}-\bm\right\}={\omega \over 2\eta_0 S}\tan(\phi/2),\hspace{.1cm}
\Im\left(\frac{1}{\ammo}\right)={k_0^3\over 6\pi\mu_0}.\l{amm_sym}
\f
This means that one should use lossless particles but not at the resonant frequency of the array, so that the real parts of the inverse polarizabilities acquire the values given in \r{aee_sym} and \r{amm_sym}.

\section{Asymmetric response to illuminations from the opposite
sides of metamirrors}

Next, we study metamirrors providing different
reflection phases when illuminated from their two sides. Equations
\r{eq:d} clearly show that in order to achieve asymmetric
full-reflection response from a single layer of small unit cells, we
need to invoke magnetoelectric coupling of the omega type
($\aemrh=\amerh\neq 0$).\footnote{Equality $\aemrh=\amerh$ follows from the
reciprocity of the structure \cite{basic}.} In order to get more physical insight into
these phenomena, as an example of asymmetric metamirrors we will
first consider the classical realization of the mushroom layer
high-impedance surface (HIS). Similarly to generic arrays of small
particles, the mushroom layer is electrically thin, and at the
resonance it has the reflection coefficient $R=+1$ (magnetic wall
response) from the side of the patch array, while the reflection coefficient from the opposite side is
$R=-1$, because the structure has a PEC ground plane. The response is
the same as we expect from an array of bi-anisotropic particles with
appropriately designed omega coupling. Furthermore, the
mushroom-layer geometry resembles that of an array of
$\Omega$-particles: An array of electrically polarizable patches
next to a PEC wall. Clearly, if this structure is excited by
external uniform electric fields, the currents induced on the patch
array and on the ground plane are different, meaning that applied
uniform electric field induces some magnetic moment (an effective
circulating electric current). What are the electric and magnetic
polarizations induced in a conventional mushroom layer and what are
the effective ``polarizabilities''? The expectation is that the
mushroom layer exhibits the same response as a resonant layer of
bi-anisotropic (omega) particles.

%%%%%%%%%%%%%%%%%%%%%%%%%%%%%%%%%%%%%
%%%%%%%%%%%%%%%%%%%%%%%%%%%%%%%%%%%%%

\subsection{Electric and magnetic polarizations in mushroom layers}
\label{HIS_polarizations}

\begin{figure}[h]
\centering
\includegraphics[width=0.45\textwidth]{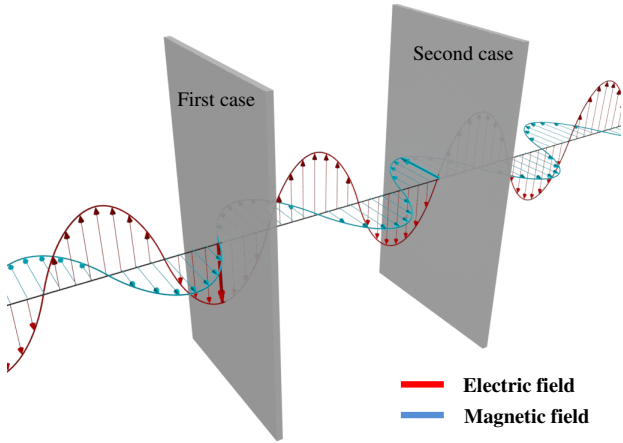}\\
  \caption{Geometry of the problem. The first case corresponds to the excitation by electric fields and the 
  second case to  the excitation by external magnetic fields.}\label{standing}
\end{figure}

%Here we show  that mushroom layer high-impedance surfaces realized
%as periodical arrays of metal patches over a PEC plane  respond to
%incident electromagnetic fields as arrays of bi-anisotropic (omega)
%particles.
To understand the electromagnetic coupling which takes
place in mushroom layers, we derive the effective polarizabilities
of these layers by studying their excitations in external electric
and magnetic fields. To find  these polarizabilities, we need to
excite the HIS layer by electric field or magnetic fields which are
approximately uniform inside the HIS layer. Following
\cite{modeboo}, we choose the exciting field in form of a standing
wave (Fig.~\ref{standing}): \e
\begin{array}{c}
\_E_{\rm inc}=\left\{E_0 e^{-jk_0z}-E_0 e^{+jk_0z}\right\}\mathbf{x}_0=-j2E_0\sin(k_0z)\mathbf{x}_0,\vspace*{.2cm}\\\displaystyle
\_H_{\rm inc}=\left\{H_0 e^{-jk_0z}+H_0 e^{+jk_0z}\right\}\mathbf{y}_0=2H_0\cos(k_0z)\mathbf{y}_0
\end{array}\l{eq:t}
\f (the Cartesian axes $\_x_0$ and $\_y_0$ are in the layer plane).

To study the excitation by external electric fields,  we position
the maximum of the electric field distribution of the standing wave
at the middle point between the patch array and the ground plane,
where $\sin(k_0z)=-1$, e.g., assuming $k_0z=-\pi/2$ (the normalized
distance between the patch array and the ground plane is $k_0d\ll 1$,
where $d$ is the distance between the ground plane and the patch
array). Next, we write the relations between the averaged surface
current density on the ground plane $J_{\rm met}$, the averaged
surface current density on the array of patches $J_{\rm g}$, and the
total surface-averaged electric fields at the respective two planes
as \e
\begin{array}{c}\displaystyle
j2E_0 \cos\left(\frac{k_0d}{2}\right)-\frac{\n}{2}J_{\rm g}-\frac{\n}{2}J_{\rm met}e^{-jk_0d}=Z_{\rm g} J_{\rm g},\vspace*{.2cm}\\\displaystyle
j2E_0 \cos\left(\frac{k_0d}{2}\right)-\frac{\n}{2}J_{\rm g}e^{-jk_0d}-\frac{\n}{2}J_{\rm met}=0,
\end{array}\l{eq:u}
\f where $Z_{\rm g}$ is the effective sheet impedance (grid
impedance) of the patch array. The value of the grid impedance we
choose so that the mushroom layer operates as a magnetic wall (the
parallel connection of $Z_{\rm g}$ and the input impedance of the
equivalent short-circuited transmission line of length $d$
$Z=j\eta_0\tan(k_0d)$ is in parallel resonance):
\e Z_{\rm
g}=-j\eta_0\tan(k_0d).\l{eq:v}\f
Using (\ref{eq:u}) and (\ref{eq:v}),
the currents on the metal surface and on the patch grid can be
found: \e
\begin{array}{c}\displaystyle
J_{\rm met}=j\frac{4}{\n}E_0 \cos\left(\frac{k_0d}{2}\right)-J_{\rm g}e^{-jk_0d},\vspace*{.2cm}\\\displaystyle
J_{\rm g}=\frac{-2E_0e^{-j\frac{k_0d}{2}}}{j\n\left(e^{-jk_0d}-\frac{1}{\cos(k_0d)}\right)}.
\end{array}\l{eq:w}
\f The induced electric moment per unit area of the structure, by
definition \cite{Definition}, is \e \displaystyle
p_x=\frac{1}{j\o}\left(J_{\rm met}+J_{\rm g}\right). \l{eq:x} \f
Using \r{eq:w} and \r{eq:x}, the electric moment can be
written as \e
\begin{array}{c}\displaystyle
\hspace*{-0.2cm}p_x=\frac{1}{j\o}\left[j\frac{4}{\n}E_0\cos\left(\frac{k_0d}{2}\right)-
\frac{2E_0e^{-j\frac{k_0d}{2}}\left(1-e^{-jk_0d}\right)}{j\n\left(e^{-jk_0d}-\frac{1}{\cos(k_0d)}\right)}\right].
\end{array}\l{eq:y}
\f
Assuming that the layer is electrically thin ($k_0d\ll 1$),  the above result simplifies to
\e
\begin{array}{c}\displaystyle
p_x=\frac{2}{\o\n}E_0.
\end{array}\l{eq:z}
\f
Dividing the induced electric moment by the amplitude of the
external  electric field $2jE_0$ in \r{eq:t}, the effective electric
polarizability per unit area is found to read
\e \displaystyle
\aeeoh=\frac{2E_0}{\o\n}\frac{1}{j2E_0}=\frac{1}{j\o\n}. \l{eq:aa}
\f
By definition \cite{Definition}, the induced magnetic moment per
unit area can be written as
\e
\begin{array}{c}\displaystyle
m_y=\frac{d\M_0}{2}\left(J_{\rm met}-J_{\rm g}\right).
\end{array}\l{eq:bb}
\f
Using \r{eq:aa} and \r{eq:bb}, the magnetic moment reads \e
\begin{array}{c}\displaystyle
\hspace*{-.1cm}m_y=\frac{d\M_0}{2}\left[j\frac{4}{\n}E_0\cos\left(\frac{k_0d}{2}\right)+\frac{2E_0e^{-j\frac{k_0d}{2}}\left(1+e^{-jk_0d}\right)}{j\n\left(e^{-jk_0d}-\frac{1}{\cos(k_0d)}\right)}\right].
\end{array}\l{eq:cc}
\f The assumption of electrically small thickness $d$ simplifies
this relation to \e
\begin{array}{c}\displaystyle
m_y=\frac{2E_0}{\o}.
\end{array}\l{eq:dd}
\f Then, using \r{eq:t} and \r{eq:dd}, we can calculate the
magnetoelectric polarizability (the omega coupling coefficient): \e
\begin{array}{c}\displaystyle
\amerh=\frac{2E_0}{\o}\frac{1}{j2E_0}=\frac{1}{j\o}.
\end{array}\l{eq:ee}
\f
At the second step, we can derive the effective  magnetic and
electromagnetic polarizabilities by considering excitation of the
same layer by external magnetic fields. To do that, we position the
layer inside the standing wave \r{eq:t} so that the maximum of the
magnetic field is between the array of patches and the ground plane
(e.g., at $k_0z=0$), see Fig.~\ref{standing}. Again writing the relations between the induced
averaged surface current densities and the total electric fields in
the plane of the patches and the ground plane \e
\begin{array}{c}\displaystyle
j2E_0 \sin\left(\frac{k_0d}{2}\right)-\frac{\n}{2}J_{\rm g}-\frac{\n}{2}J_{\rm met}e^{-jk_0d}=Z_{\rm g} J_{\rm g},\vspace*{.2cm}\\\displaystyle
-j2E_0 \sin\left(\frac{k_0d}{2}\right)-\frac{\n}{2}J_{\rm g}e^{-jk_0d}-\frac{\n}{2}J_{\rm met}=0,
\end{array}\l{eq:ff}
\f
the current densities on the metal surface and on the grid array are found as
\e
\begin{array}{c}\displaystyle
J_{\rm met}=-j\frac{4}{\n}E_0 \sin\left(\frac{k_0d}{2}\right)-J_{\rm g}e^{-jk_0d},\vspace*{.2cm}\\\displaystyle
J_{\rm g}=\frac{j2E_0e^{-jk_0\frac{d}{2}}}{j\n\left(e^{-jk_0d}-\frac{1}{\cos(k_0d)}\right)}.
\end{array}\l{eq:gg}
\f
The electric moment is calculated using \r{eq:x}:
\e
\begin{array}{c}\displaystyle
p_x=\frac{1}{j\o}\left[\frac{j2E_0e^{-jk_0d}\left(1-e^{-jk_0d}\right)}{j\n\left(e^{-jk_0d}-\frac{1}{\cos(k_0d)}\right)}\right].
\end{array}\l{eq:hh}
\f
Assuming electrically small thickness for the structure, we get
\e
\begin{array}{c}\displaystyle
p_x=-\frac{2}{j\o}H_0.
\end{array}\l{eq:ii}
\f
The coupling coefficient can be found dividing the induced electric moment by the amplitude of the external magnetic field in \r{eq:t}
\e
\begin{array}{c}\displaystyle
\aemrh=-\left(\frac{-2H_0}{j\o}\right)\frac{1}{2H_0}=\frac{1}{j\o}.
\end{array}\l{eq:jj}
\f
Using \r{eq:bb}, the magnetic moment can be written as
\e
\begin{array}{c}\displaystyle
m_y=\frac{d\M_0}{2}\left[\frac{-j2E_0e^{-j\frac{k_0d}{2}}\left(1+e^{-jk_0d}\right)}{j\n\left(e^{-jk_0d}-\frac{1}{\cos(k_0d)}\right)}\right],
\end{array}\l{eq:kk}
\f
and within the same assumption of electrically small thickness,
\e
\begin{array}{c}\displaystyle
m_y=\frac{2E_0}{j\o}.
\end{array}\l{eq:ll}
\f
The magnetic polarizability can be determined by dividing the induced magnetic moment by the external magnetic field in \r{eq:t}:
\e
\begin{array}{c}\displaystyle
\ammoh=\frac{2\n H_0}{j\o}\frac{1}{2H_0}=\frac{\n}{j\o}.
\end{array}\l{eq:mm}
\f

\subsection{Analogy between mushroom layers and arrays of resonant
omega particles}

The  results for the polarizabilities \r{eq:ee} and \r{eq:jj} show
that the mushroom layer at resonance indeed acts as a layer of omega
particles, because the structure develops electric polarization at
excitation by electric fields, magnetic polarization under
excitation by magnetic fields, and, in addition,  there is
magneto-electric coupling effect measured by the coefficients \e
\begin{array}{c}\displaystyle
\aemrh=\amerh ={1\over j\omega}.
\end{array}\l{eq:nn}
\f Furthermore, we note that the polarizabilities are balanced as \e
\n \aeeoh=\aemrh=\amerh={1\over \n} \ammoh={1\over j\omega}.
\l{eq:oo}\f According to \r{eq:d},  this is exactly the condition
ensuring that the reflection coefficient from the array of
bi-anisotropic particles changes sign when we illuminate the array
from the opposite side. Moreover, we recall that the
polarizabilities found for the mushroom layer are defined in terms
of induced moments per unit area, which are related to the
polarizabillities of particles (or unit cells) in \r{eq:b} as
$\alpha_{ij}\rightarrow \alpha_{ij}/S$. With this in view, we see
that the amplitudes of the cross polarizabilities of the mushroom
layer are exactly what are required to realize $R=\pm 1$ response of
an array of omega particles. This consideration again confirms that in order
to realize a single-layer full-reflection structure with asymmetric
phase responses from different sides, we have to introduce
bi-anisotropy inside the layer. Next, we will derive  design
equations for an ultimately thin full-reflection layer which has
arbitrary (asymmetric) reflection phases for waves coming from the
two opposite sides.

\section{General metamirrors}

Let us find the required polarizabilities of unit cells for
realization of general asymmetric metamirrors. These layers fully
reflect plane waves coming from both sides (the amplitude of the
reflection coefficient is equal to unity), but  the phases of the
reflected waves can take any desired values. PEC, perfect magnetic
conductor (PMC), HIS, and layers with $\pm j$ reflection
coefficients will be considered as important special cases. Let us
demand that the reflection coefficients equal to $e^{j\phi}$ and
$e^{j\theta}$ for $-\mathbf{z}_0$ and $+\mathbf{z}_0$-directed
incident plane waves, respectively: \e
R_{-\mathbf{z}_0}=e^{j\phi},\quad R_{+\mathbf{z}_0}=e^{j\theta}.
\l{eq:pp}\f The requirements for the effective polarizabilities of unit cells necessary to realize the desired response can be found,
substituting \r{eq:pp} in \r{eq:d}, as \e
\begin{array}{c}\displaystyle
\n\aeeoh=\frac{S}{j\o}\left[1-\frac{e^{j\phi}+e^{j\theta}}{2}\right],\vspace*{0.2cm}\\\displaystyle
\aemrh=\amerh=\frac{-S}{j\o}\left[\frac{e^{j\phi}-e^{j\theta}}{2}\right],\vspace*{0.2cm}\\\displaystyle
\frac{1}{\n}\ammoh=\frac{S}{j\o}\left[1+\frac{e^{j\phi}+e^{j\theta}}{2}\right].
\end{array}\l{eq:qq}\f
The polarizabilities in \r{eq:qq} refer to a layer with omega coupling.
This is physically understandable, because we demand this layer to
show the same transmission properties but different reflection
properties for waves incident from the opposite directions, while
the polarization of the reflected and transmitted waves is the same as
that of the incident waves.

To find the required individual polarizabilities of single inclusions (not interacting with other inclusions), we can use the explicit formulas for the collective polarizabilities of an omega particle in an array in terms of the individual polarizabilities and the interaction constants \cite{Teemu}
\e
\begin{array}{c}
\displaystyle
\hspace{-.2cm}\aeeo=\frac{\aeeoh+\bm(\aeeoh\ammoh+\aemrh\amerh)}{1+(\aeeoh\be+\ammoh\bm)+\be\bm(\aeeoh\ammoh+\aemrh\amerh)},
\vspace*{.2cm}\\\displaystyle
\hspace{-.2cm}\ammo=\frac{\ammoh+\be(\aeeoh\ammoh+\aemrh\amerh)}{1+(\aeeoh\be+\ammoh\bm)+\be\bm(\aeeoh\ammoh+\aemrh\amerh)},
\vspace*{.2cm}\\\displaystyle
\hspace{-.2cm}\aemr=\frac{\aemrh}
{1+(\aeeoh\be+\ammoh\bm)+\be\bm(\aeeoh\ammoh+\aemrh\amerh)}.
\end{array}\l{eq:rr}
\f
Here $\aemr=\amer$ is the omega coupling parameter of an individual inclusion. Using \r{eq:qq} and \r{eq:rr}, one can find the required polarizability values for the individual omega particles as
\e
\begin{array}{c}
\displaystyle
\n\aeeo=\frac{1-e^{j(\theta+\phi)}+\frac{j\o\n}{\be S}\left[-\frac{1}{2}\left(e^{j\phi}+e^{j\theta}\right)+1\right]}{-e^{j(\theta+\phi)}+\left(1+j\frac{\o\n}{\be S}\right)^2}\frac{\n}{\be},
\vspace*{.2cm}\\\displaystyle
\aemr=\amer=\frac{-\frac{j\o\n}{2\be S}\left(e^{j\phi}-e^{j\theta}\right)}{-e^{j(\theta+\phi)}+\left(1+j\frac{\o\n}{\be S}\right)^2}\frac{\n}{\be},
\vspace*{.2cm}\\\displaystyle
\frac{1}{\n}\ammo=\frac{1-e^{j(\theta+\phi)}+\frac{j\o\n}{\be S}\left[\frac{1}{2}\left(e^{j\phi}+e^{j\theta}\right)+1\right]}{-e^{j(\theta+\phi)}+\left(1+j\frac{\o\n}{\be S}\right)^2}\frac{\n}{\be}.
\end{array}\l{eq:ss}
\f
%From \r{eq:qq} and \r{eq:ss}, it is clear that if we want to have asymmetry in reflection from different sides $\theta\neq\phi$, there must be omega coupling introduced in the particles.
These individual polarizabilities satisfy the necessary condition for  lossless bi-anisotropic particles \cite{double_array}
\e
\begin{array}{c}
\displaystyle
\Im\left\{\left(\aeeb-\aemb\.\ammb^{-1}\.\ameb\right)^{-1}\right\}=\vspace*{0.2cm}\\\displaystyle
\Im\left\{\be+\displaystyle\frac{j\o\n}{2S}\frac{e^{j\phi}+e^{j\theta}+2}{1-e^{j(\phi+\theta)}}\right\}\It=\displaystyle\frac{k_0^3}{6\pi\E_0}\It.
\end{array}\l{eq:tt}
\f
These results are quite general in the sense that we can design a layer of bi-anisotropic particles which fully reflects coming electromagnetic waves from different sides with different phases.

\begin{figure}[h]
\centering
\includegraphics[width=.8\columnwidth]{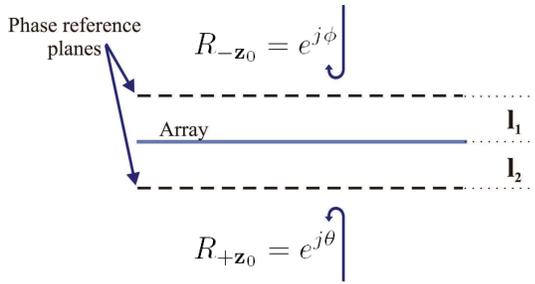}
\caption{Required reflection coefficients at distances $l_1$ and $l_2$ from the layer.}
\label{fig:Distance}
\end{figure}

For some applications it can be necessary to ensure a specific value of the phase of the reflected wave at some specific distance from the layer plane. Our theory can be easily modified to realize any phase requirements at any distance from the array. Suppose that we want to ensure that the phase of the reflected wave equals $\phi$ at distance $l_1$ from the upper surface of the array and phase $\theta$ for reflected waves at distance $l_2$ from the bottom surface of the array (see Fig.~\ref{fig:Distance}). To find the required effective and individual polarizabilities, we can simply substitute $\phi\rightarrow\phi+4\pi l_1/\lambda $ and $\theta\rightarrow\theta+4\pi l_2/\lambda $
in \r{eq:rr} and \r{eq:ss}, where $\lambda$ is the free-space wavelength.

\begin{figure}[h]
  \centering
  \hfill\subcaptionbox{\label{fig:1a}}{\centering \includegraphics[width=0.45\textwidth]{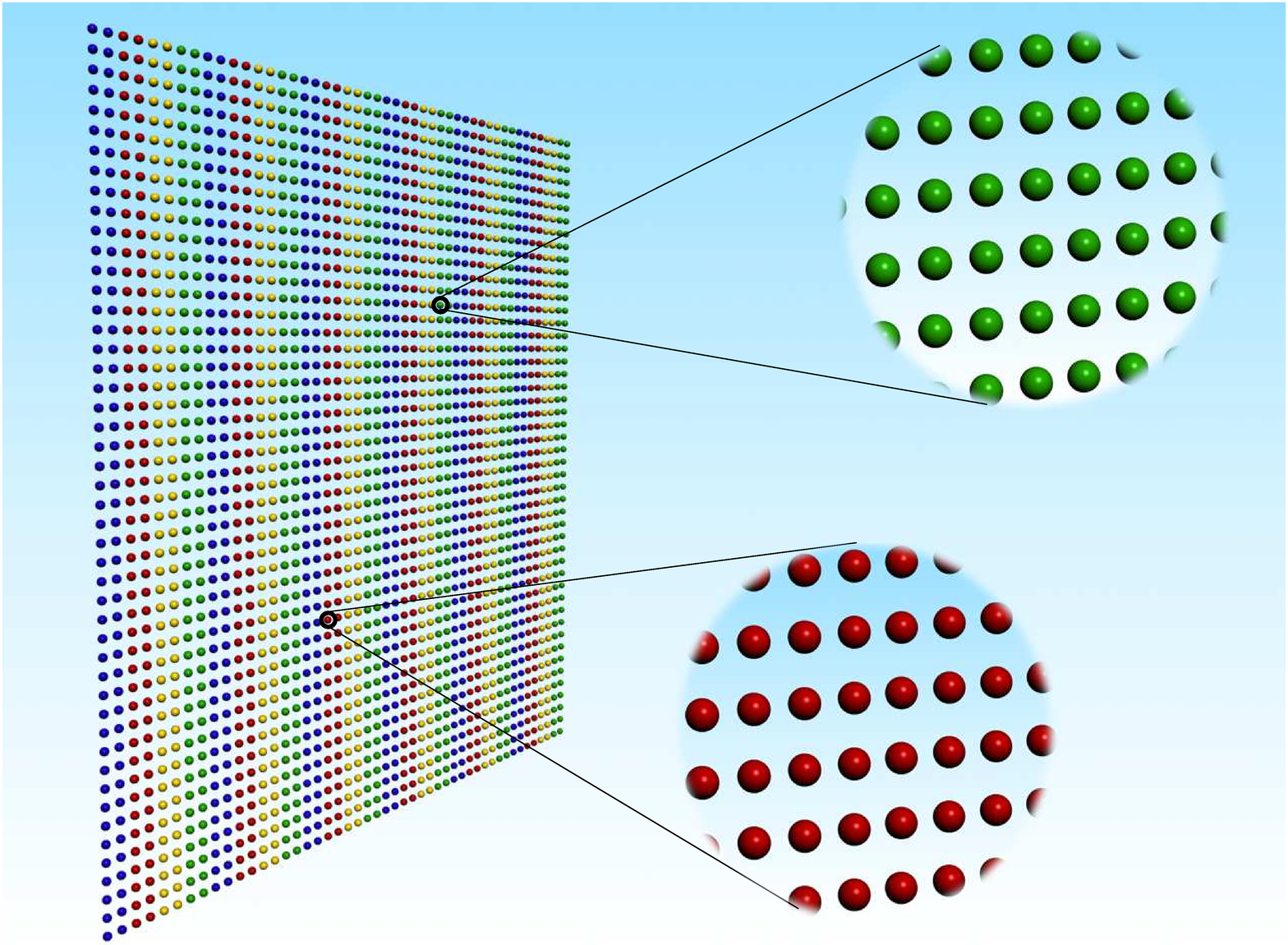}} \\
  \hfill\subcaptionbox{\label{fig:1b}}{\centering \includegraphics[width=0.45\textwidth]{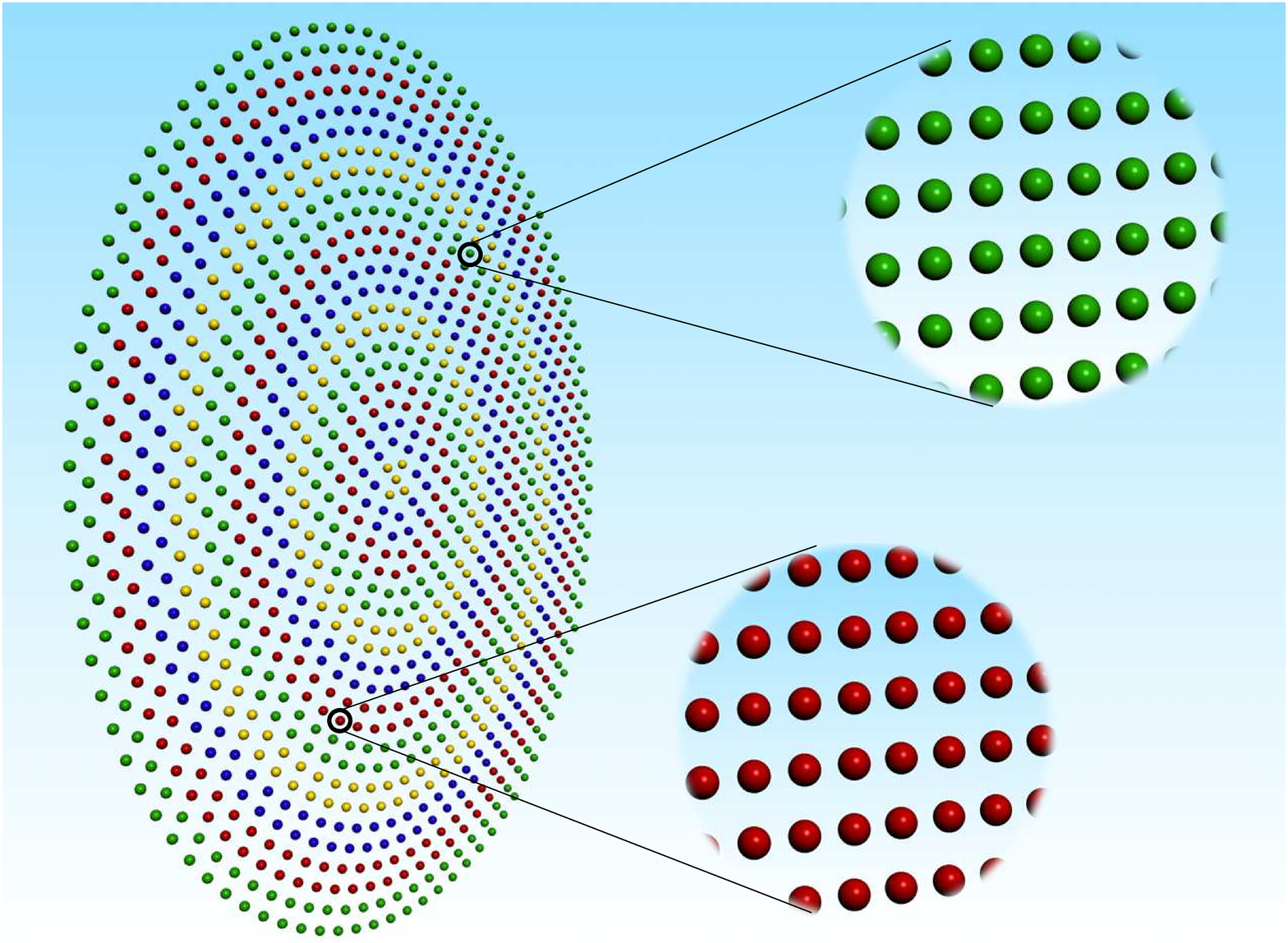}}
  \caption{Examples of dipolar arrays for tailoring reflected waves. (a) Changing the direction of the reflected wave. (b) Realizing a reflectarray using a single array of particles.}
  \label{Single}
\end{figure}

The results in \r{eq:ss} have been derived for periodical arrays at normal incidence, meaning that we have uniform phase distribution in reflected waves from both sides of the surface. Obviously, for many applications it is desirable to realize specific non-uniform phase distributions of reflected plane waves (allowing, for instance, wave focusing or deflection). Moreover, it is of practical interest to understand if it is possible to realize \emph{different phase distributions} for illuminations from the opposite sides, using just one single layer of dipolar particles.  In the assumption that the variations of the reflection phase along the surface is slow and smooth on the wavelength scale, the physical optics approximation \cite{Kirchhoff} can be used to determine the required individual properties for each particle at each point on the surface (see an illustration in Fig.~\ref{Single}). Knowing the required  reflection phase value at each point of the surface, one can replace each particle on the array locally by a periodic array formed by many such particles. This way the necessary collective and consequently individual polarizabilities of the particle can be calculated from \r{eq:qq} and \r{eq:ss}, respectively. An interesting conclusion is that it is possible to realize a single layer of bi-anisotropic particles which offers different reflection phase distributions from two sides. For example, a layer can exhibit uniform reflection phase distribution when illuminated from one side,  and a non-uniform reflection phase distribution from the other side. One interesting example is an ultimately thin layer of dipolar particles which works as a reflectarray from one side and as a magnetic wall from the other side.

Interestingly, the full control over reflection phases of waves reflected from both sides of metamirrors can be achieved using dipolar particles of only one type: lossless omega particles.
There is an important point to be noticed. The amplitudes of the reflection and transmission coefficients equal unity and zero, respectively, and the structure is obviously lossless. However, this implies only that the particles are lossless in the sense of the overall response (the power extracted from the fields equals to the re-radiated power). Because the electric and magnetic responses are coupled in these bi-anisotropic particles, it is possible, for instance, that  the electric and magnetic polarizabilities provide gain, which is compensated by equivalent loss in the omega coupling term.

In what follows, we will consider special cases of this general case. The cases of PEC and PMC response have been already treated above (Section~\ref{PEC_PMC}), and we will concentrate on high-impedance surfaces and surfaces providing the reflection coefficients equal to $\pm j$.

\subsection{Special cases of metamirrors}

\subsubsection{$\pm j$ reflection coefficients}
Let us consider metamirrors with the reflection coefficients equal to
 $R_{\mp\mathbf{z}_0}=\pm j$ when illuminated from their two sides. Note that the input impedance of such a layer seen at its two surfaces is $Z_{\rm in}^{\mp\mathbf{z}_0}=\pm j\n$. Thus, realizing these layers is the same as designing a  thin structure whose input impedance has a $90^\circ$ phase shift from the intrinsic impedance of the background medium. Such a layer might find important applications in antenna techniques, for example.

Substituting $\phi=-\theta=\frac{\pi}{2}$ in \r{eq:pp} and then in \r{eq:qq}, we can determine the required effective polarizabilities as follows:
\e
\begin{array}{c}\displaystyle
\n\aeeoh=j\aemrh=\frac{1}{\n}\ammoh=\frac{S}{j\o}.
\end{array}\l{eq:bbb}\f
As it is seen, the effective polarizabilities have the balanced properties (the magnitudes of all the normalized polarizabilities are equal). Using \r{eq:ss}, the individual polarizabilities can be written as
\e
\begin{array}{c}
\displaystyle
\n\aeeo=j\aemr=\frac{1}{\n}\ammo=\frac{1}{2+j\frac{\o\n}{\be S}}\frac{\n}{\be}.
\end{array}\l{eq:ccc}
\f
It should be noted that in this case both effective and individual polarizabilities have the balanced strength.

\subsubsection{High-impedance surface}
One of the practically important cases is a single layer of bi-anisotropic particles acting as a high-impedance surface. In Section~\ref{HIS_polarizations} it was shown that there is omega coupling inside conventional HIS mushroom layers. Here, we find the required effective and individual polarizabilities of dipolar bi-anisotropic particles which realize the same response. To ensure HIS properties of an array of bi-anisotropic particles, we demand that $\phi=\pi$ and  $\theta=0$ in \r{eq:pp}. Substituting these values  into \r{eq:qq}, one can get the required effective polarizabilities as
\e
\begin{array}{c}\displaystyle
\n\aeeoh=\aemrh=\frac{1}{\n}\ammoh=\frac{S}{j\o}.
\end{array}\l{eq:yy}\f
As expected, these are the same relations as we have found for the effective polarizabilities (defined per unit area) of conventional mushroom layers \r{eq:oo}.
Using \r{eq:ss}, the individual polarizabilities can be found as
\e
\begin{array}{c}
\displaystyle
\n\aeeo=\frac{1}{\n}\ammo=\frac{2+j \frac{\o\n}{\be S}}{1+(1+j\frac{\o\n}{\be S})^2}\frac{\n}{\be},
\vspace*{.2cm}\\\displaystyle
\aemr=\amer=\frac{-j\frac{\o\n}{\be S}}{1+(1+j\frac{\o\n}{\be S})^2}\frac{\n}{\be}.
\end{array}\l{eq:zz}
\f
Interestingly, the effective polarizabilities are balanced, but the individual ones are not completely balanced.

%However, it can be shown that the polarizabilities in \r{eq:zz} are quite close to balanced case in which all normalized polarizabilities have the same amplitude
%\e
%\begin{array}{c}
%\displaystyle
%\n\aeeo=-j\aemr=\frac{1}{\n}\ammo.
%\end{array}\l{eq:aaa}
%\f
%It will be shown that this small imbalanced properties between coupling and other polarizabilities will make huge difference in operation from the case of purely balanced particle.

\section{Example designs}
The above theory determines the required polarizabilities of unit cells which ensure the desired properties of general full-reflection layers. Here we will consider example designs of particles which such polarizabilities. It is clear that to realize PEC and PMC responses from arrays of small dipole particles, we can simply use periodical array of resonant and low-loss electric and magnetic dipoles, respectively. But what particles can have the required polarizabilities as in \r{eq:ccc} and \r{eq:zz}? In what follows, we will focus on asymmetric responses from different sides of the array.

\subsection{Designing a layer with $\pm j$ reflection coefficients}
\begin{figure}[h]
\centering
\begin{subfigure}{0.5\columnwidth}
  \centering
  \includegraphics[width=\columnwidth]{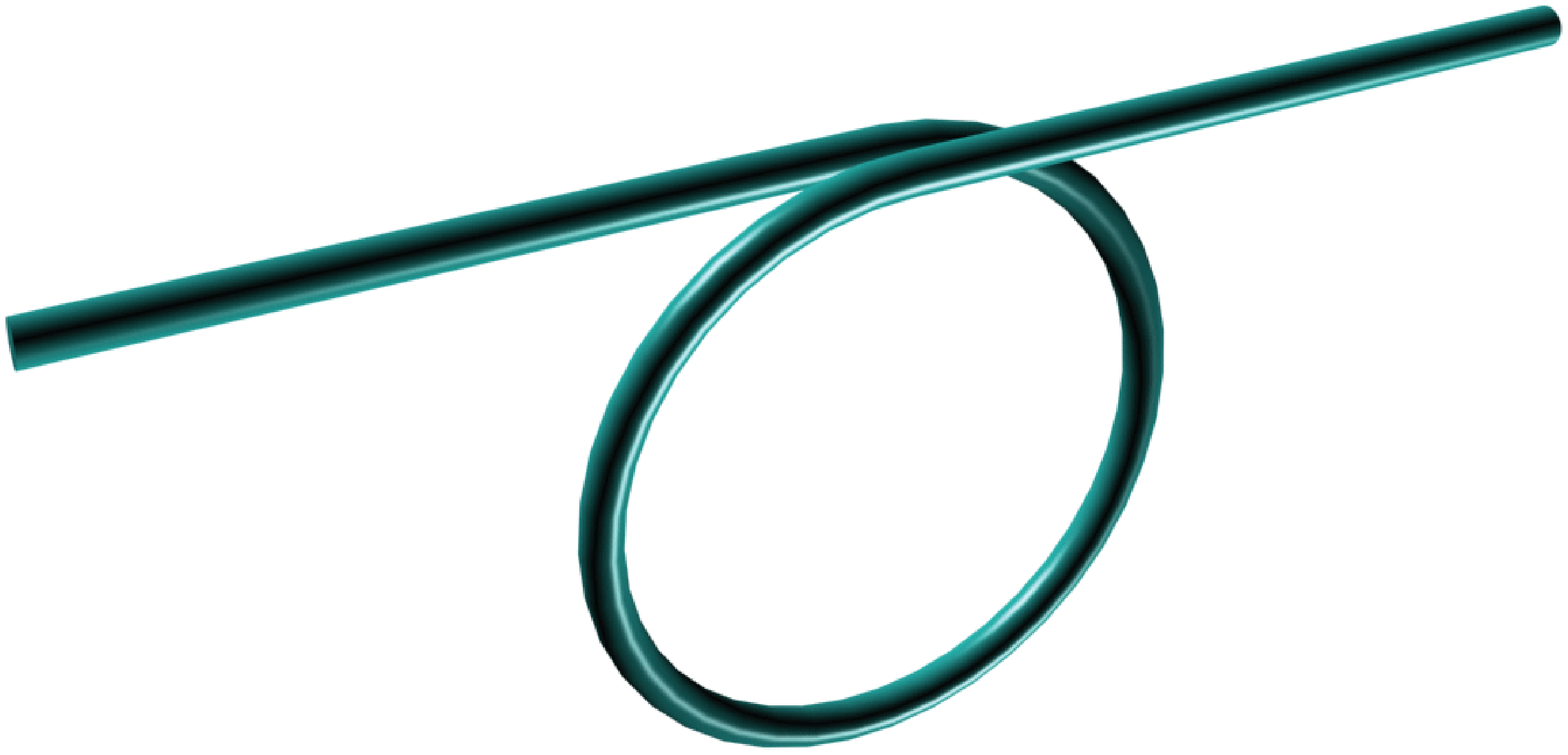}
  \caption{}
  \label{fig:LOP}
\end{subfigure}%
\begin{subfigure}{0.5\columnwidth}
  \centering
  \includegraphics[width=\columnwidth]{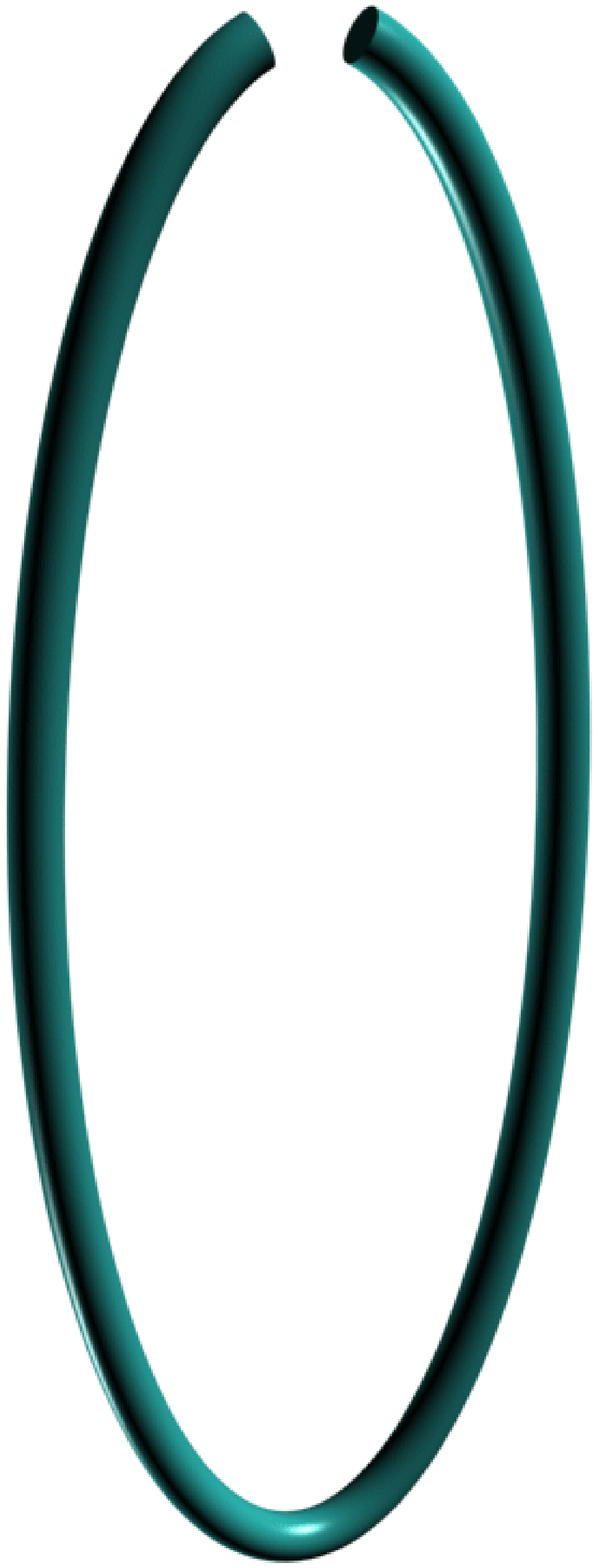}
  \caption{}
  \label{fig:ROP}
\end{subfigure}
\caption{(a) Conventional omega particle. (b) Ellipsoidal omega particle. Uniaxial (rotationally symmetric) structures are formed by pairs of two particles.}
\label{fig:test}
\end{figure}

\begin{figure}[h]
\begin{minipage}{\columnwidth}
\centering
\includegraphics[width=\columnwidth]{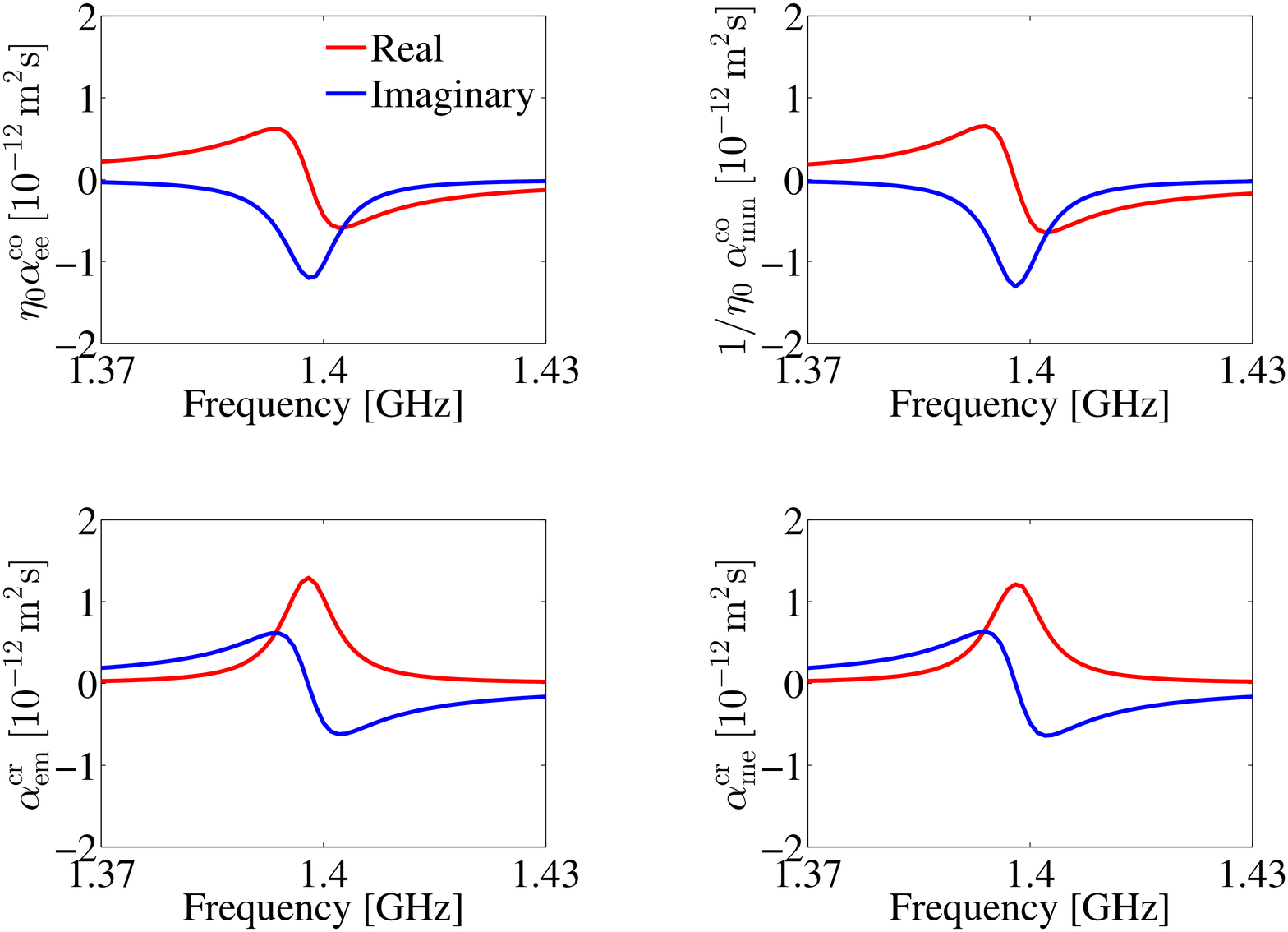}
\subcaption{}
\label{fig:LOP-P}
\end{minipage}
\begin{minipage}{\columnwidth}
\centering
\includegraphics[width=\columnwidth]{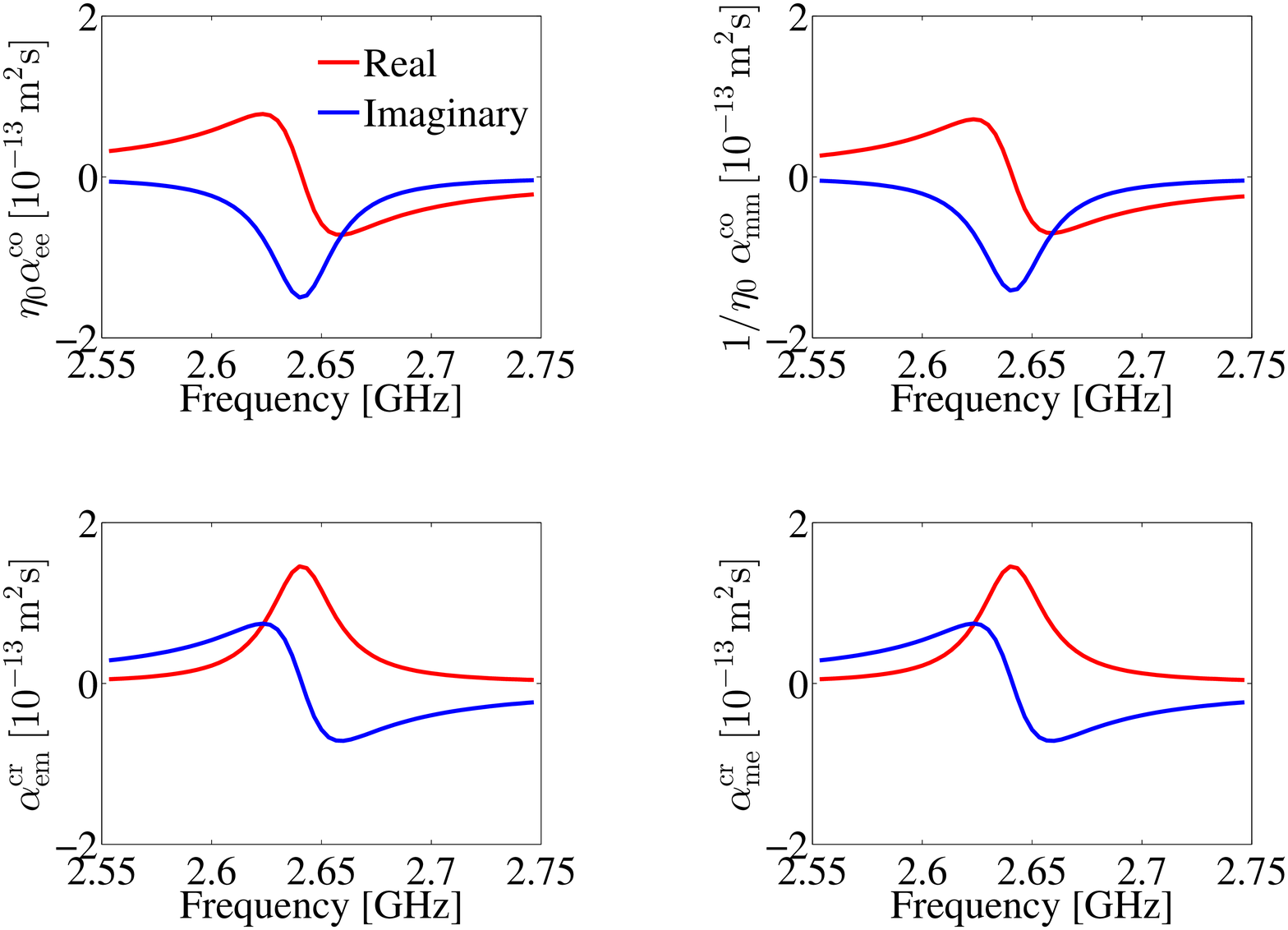}
\subcaption{}
\label{fig:ROP-P}
\end{minipage}
\caption{Individual polarizabilities for: (a) Conventional omega particle. (b) Ellipsoidal omega particle.}
\label{two figures stacked}
\end{figure}

\begin{figure}[h]
\begin{minipage}{\columnwidth}
\centering
\includegraphics[width=\columnwidth]{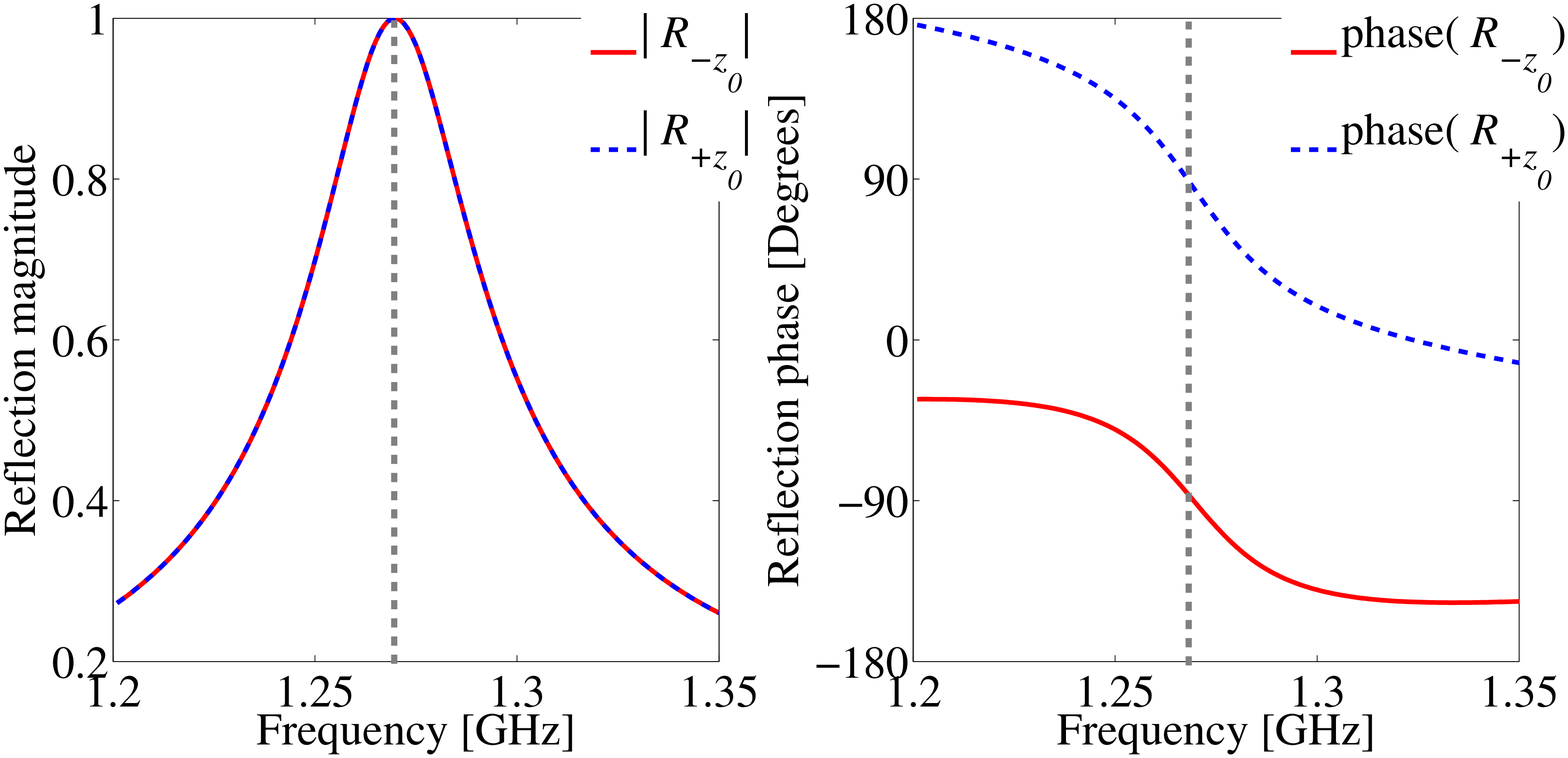}
\subcaption{}
\label{fig:LOP-CR}
\end{minipage}
\begin{minipage}{\columnwidth}
\centering
\includegraphics[width=\columnwidth]{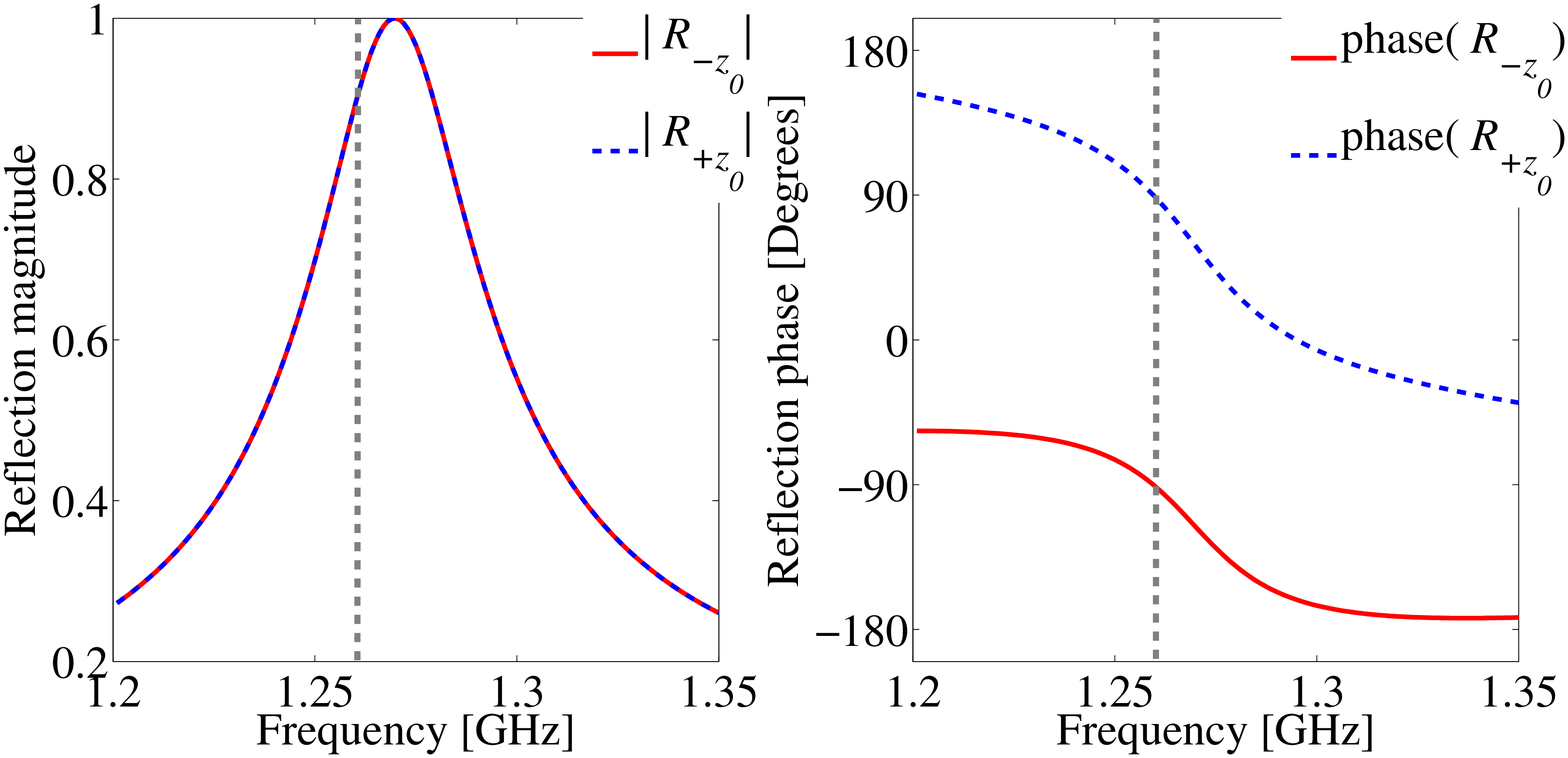}
\subcaption{}
\label{fig:LOP-BR}
\end{minipage}
\caption{(a) Realizing $R_{\pm\mathbf{z}_0}=\pm j$ reflection coefficients from an array made of conventional omega particles when the phase reference plane is set to the center of the array. (b)  Realizing $R_{\pm\mathbf{z}_0}=\pm j$ reflection coefficients from an array made of conventional omega particles when the phase reference plane is set to the borders of the array.}
\label{fig:OP-R}
\end{figure}

In the previous section, it was shown that to realize a layer which exhibits $R_{\pm\mathbf{z}_0}=\pm j$ reflection coefficient for waves coming from different sides of the array, one needs to have an array of balanced omega particles. In this section, as examples of omega particles for microwave applications, we will study metal-wire omega particles. Figs.~\ref{fig:LOP} and \ref{fig:ROP} show two possible designs of bi-anisotropic particles possessing omega coupling.
To optimize the particle shapes and dimensions, we use the method introduced in \cite{Viitya} which allows us to extract the polarizabilities of an arbitrary dipolar
particle from computed or measured scattering cross sections of the particle. The individual polarizabilities of particles optimized for the desired performance are shown in Figs.~\ref{fig:LOP-P} and \ref{fig:ROP-P}.
As it is seen, both of these particles are balanced and can be used to realize a layer with $R_{\pm\mathbf{z}_0}=\pm j$ reflection. Here, we utilize the particle shown in  Fig.~\ref{fig:LOP}. The radius of the loop is $r=7.45$~mm, the radius of the wire is $r_0=0.5$~mm, the half-length of the electric dipole is $l=18.1$~mm, and the pitch is $1.45$~mm. Fig.~\ref{fig:LOP-CR} shows the amplitude and phase of reflected waves from an array made of these particles (the array period is $a=44$~mm).  $R_{\pm\mathbf{z}_0}=\pm j$ reflection coefficients are realized at the particle resonance (the reference phase plane crosses the geometrical center of the particles).

Let us next consider the case when it is required to realize the reflection coefficients equal to  $+j$ at distance $l_1$ from the top of the array and $-j$  at distance $l_2$ from the bottom of the array (see Fig.~\ref{fig:Distance}). As it was explained in the previous section, the present theory can be modified to realize any reflection phase requirement at any distance from the array. To do this, one can substitute $\phi=\pi/2(1+8/\lambda l_1)$ and $\theta=-\pi/2(1-8/\lambda l_2)$ in \r{eq:ss}. As it is clear from \r{eq:ss}, in this case the required polarizabilities are not balanced. For example, the particle in Fig.~\ref{fig:LOP} can be optimized to get a layer with $R_{\pm\mathbf{z}_0}=\pm j$ reflection coefficients from the borders of the particle (when the phase reference planes are set to the borders of the particle). In this case, the radius of the loop is $r=7.6$~mm and the half-length of the electric dipole is $l=17.6$~mm. As we see from Fig.~\ref{fig:LOP-BR}, at the frequency where $R_{\pm\mathbf{z}_0}=\pm j$ is realized, the amplitude of the reflection coefficient is not exactly equal to unity. On the other hand, at the frequency where we realize full reflection, the phases of the reflection coefficients from different sides are not exactly equal to $\pm \pi/2$. The reason is the intrinsic relation between electric, magnetic and magnetoelectric polarizabilities of small wire omega particles  (see \cite{basic})
\e
\aeeo\ammo=-\aemr\amer.
\l{eq:ooo}\f
In some cases, this condition limits the accessible properties of wire omega particles which does not allow one to fully satisfy the requirements in \r{eq:ss}. However, as is shown in \cite{relation}, this restriction is due to specific symmetry properties of induced current distribution and does not hold for general omega particles.

\subsection{Designing a layer with $R_{\pm\mathbf{z}_0}=\pm 1$ reflection coefficients}

It was shown above that  to design a layer of bi-anisotropic particles which acts as a HIS one needs particles with balanced electric and magnetic polarizabilities, but the magnetoelectric coupling coefficient should not be balanced with the electric and magnetic polarizabilities (see \r{eq:zz}). It is seen from \r{eq:ooo} that there is an intrinsic limitation for wire omega particles. Condition \r{eq:ooo} shows that as soon as we have a wire omega particle with balanced electric and magnetic polarizabilities, the coupling coefficient will be automatically balanced with the electric and magnetic ones. This is in contrast with the requirement in \r{eq:zz}. Let us still consider another possibility to use wire omega particles to realize $R_{\pm\mathbf{z}_0}=\pm 1$ reflection layer, which appears thanks to the freedom to choose the area of the unit cells $S$ so that all the polarizabilities in \r{eq:zz} become balanced. Substituting the interaction constants from \r{eq:j}, the required polarizabilities in \r{eq:zz} can be rewritten as
\e
\begin{array}{c}
\displaystyle
\n\aeeo=\frac{1}{\n}\ammo=\n\frac{\Re\{\be\}+j \frac{k_0^3}{6\pi\epsilon_0}}{(\Re\{\be\}+j\frac{k_0^3}{6\pi\epsilon_0})^2-\left(\frac{\o\n}{2 S}\right)^2},
\vspace*{.2cm}\\\displaystyle
\aemr=\amer=\n\frac{-\frac{\o\n}{2 S}}{(\Re\{\be\}+j\frac{k_0^3}{6\pi\epsilon_0})^2-\left(\frac{\o\n}{2 S}\right)^2}.
\end{array}\l{eq:ppp}
\f
However, as it is seen from these requirements, the only solution when all the polarizabilities are balanced is for the infinite polarizabilities, which is not realizable. This means that the conventional wire omega particle is not the right choice for the design of such layers and we need an omega particle which does not obey the limitation \r{eq:ooo}.

\begin{figure}[h]
\begin{minipage}{\columnwidth}
\centering
\includegraphics[width=\columnwidth]{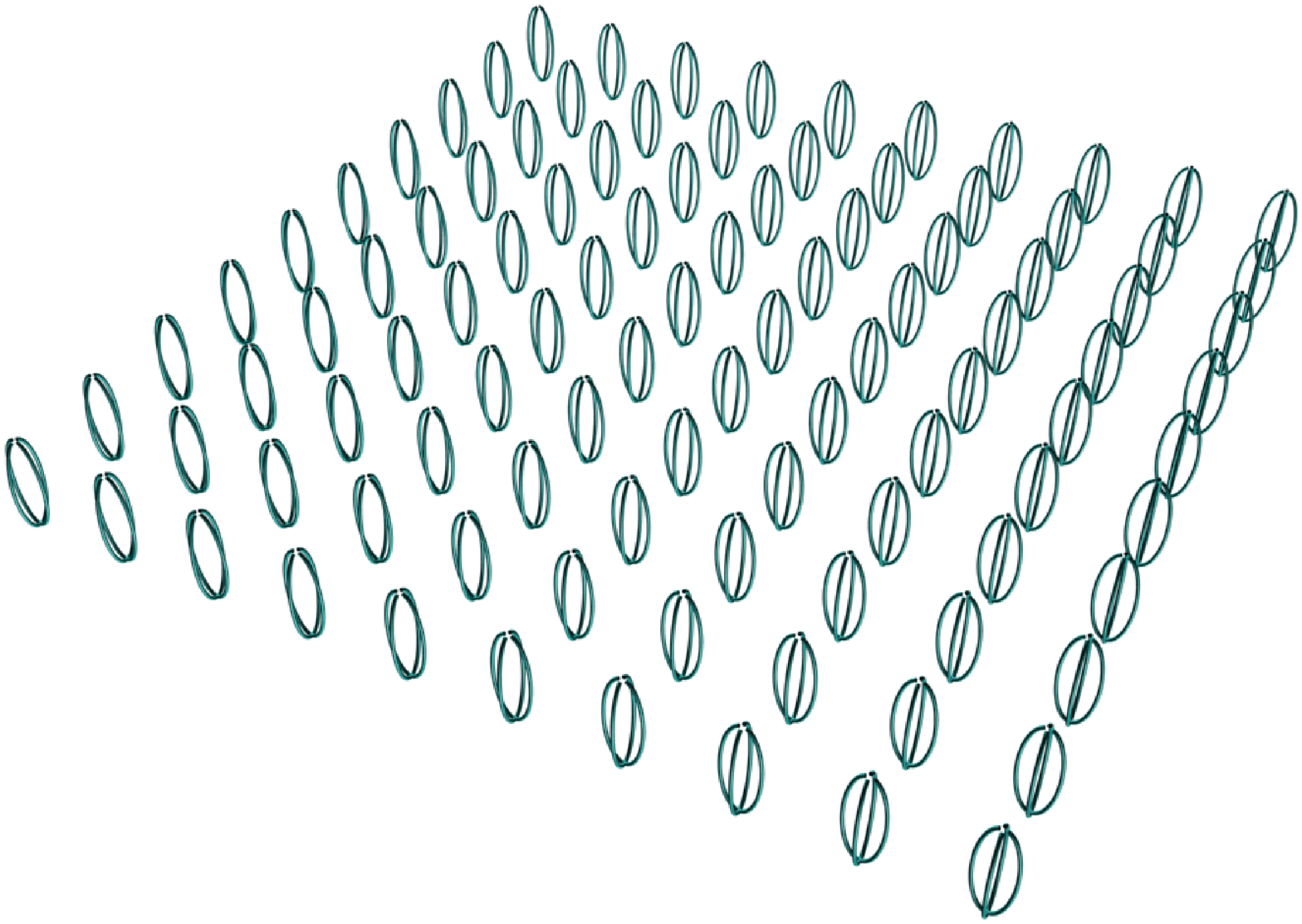}
\subcaption{}
\label{fig:Array}
\end{minipage}
\begin{minipage}{\columnwidth}
\centering
\includegraphics[width=\columnwidth]{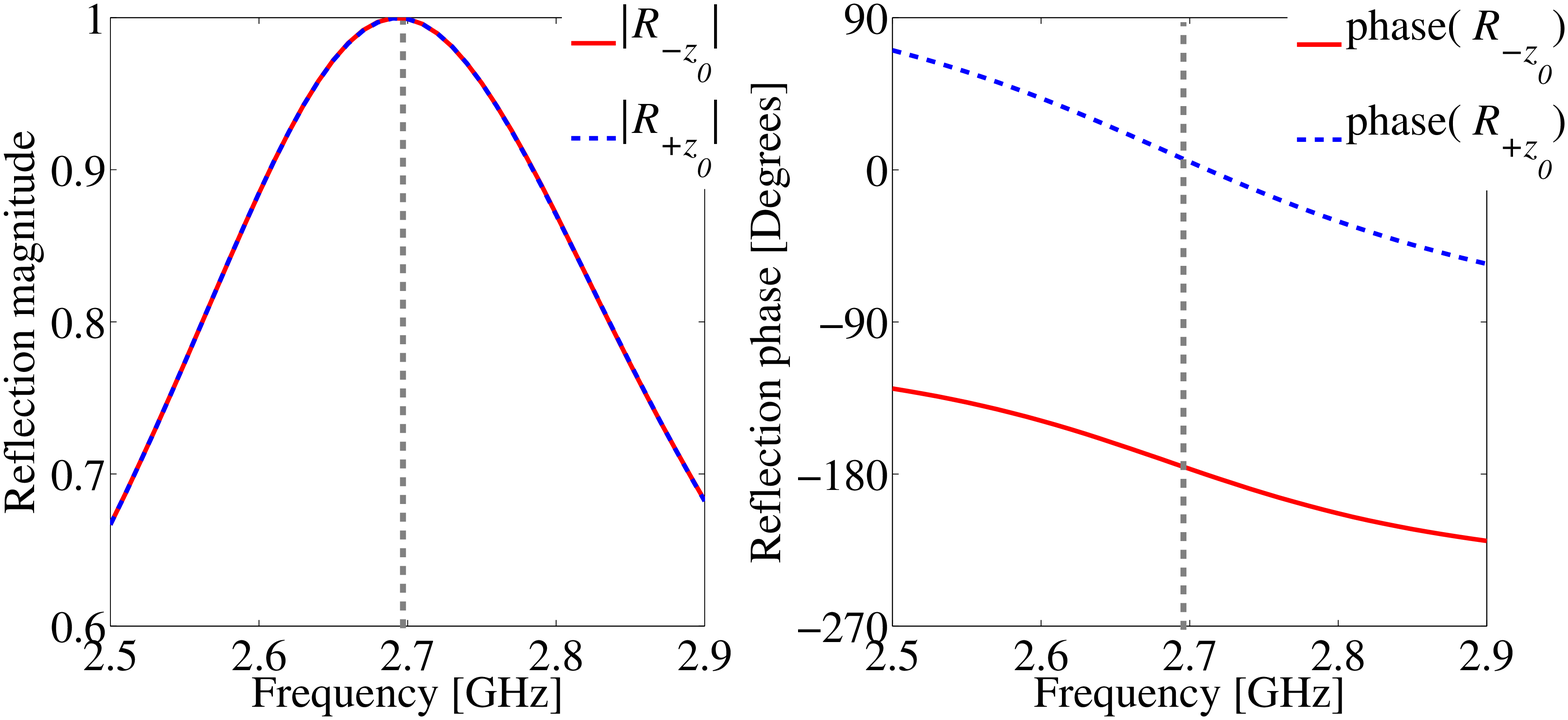}
\subcaption{}
\label{fig:ROP-BR}
\end{minipage}
\caption{(a) An array of ellipsoidal omega particles. (c) $R_{\pm\mathbf{z}_0}=\pm 1$ reflection coefficients from an array made of ellipsoidal omega particles when the phase reference plane is set to the border surfaces of the array.}
\label{fig:BR}
\end{figure}

From the applications point of view, it is better to have a layer in which we have $R_{\pm\mathbf{z}_0}=\pm 1$ reflection coefficients at the border surfaces of the layer. Fig.~\ref{fig:ROP} shows another topology of an omega particle. After optimizing the particle, the radius of the wire is $r_0=0.5$~mm, and the ellipticity is $0.334$. The individual polarizabilities of this optimized particle plotted in Fig.~\ref{fig:ROP-P} show that this particle is indeed balanced, meaning that if we set the phase reference plane at the center of the array,  $R_{\pm\mathbf{z}_0}=\pm j$ reflection from the two sides will be achieved. For an array made of these particles (the array period is $a=16$~mm) shown in Fig.~\ref{fig:Array}, the amplitude and phase of the reflection coefficients measured at the border surfaces of the array are plotted in Fig.~\ref{fig:ROP-BR}. As we see, because of the special geometry of the particle, we are able to realize $R_{\pm\mathbf{z}_0}=\pm 1$ reflection coefficients at the border surfaces of the array.

It should be noted that the considered omega particles are just very simple and particular examples which were used to show the possibility of realizing different kinds of composite metamirrors. Actually, in these examples the layer of these particles is not very thin electrically. The thicknesses of the presented example arrays of conventional and ellipsoidal omega particles are $\lambda/15.5$ and $\lambda/4.3$, respectively. However, one can use any other particle which has omega coupling to realize the required polarizabilities in much smaller  volumes. It is interesting to note that the thin focusing layer presented in \cite{Bozhevolnyi} actually has the appropriate symmetry of the omega-type bi-anisotropic layer, although without theoretical optimization and with high losses in metal shows only moderate reflectivity.

\section{Conclusions}

Here, we have introduced the concept of the metamirror which offers a possibility for full control over the reflection phases of its two surfaces. We have assumed that the metamirror thickness is ultimately small, only allowing one layer of dipolar particles to fit inside the structure. The investigation of the necessary polarization properties of the unit cells of the structure has shown that reciprocal and lossless bi-anisotropic particles with omega magnetoelectric coupling are the appropriate building blocks for the most general metamirrors. The study has also revealed the bi-anisotropic nature of the electromagnetic response of conventional high-impedance surfaces in form of mushroom layers (these layers provide the reflection coefficients equal to $+1$ and $-1$ when illuminated from the opposite sides --- a very special case of the general metamirror response). The bi-anisotropic coupling required  for the metamirror operation can be realized by selecting an appropriate shape of dipolar particles sitting in each unit cell of the planar metamirror array. For the microwave frequency range metal wire particles can be used, and we have demonstrated by simulations the metamirror responses of some example designs. For infrared and optical frequencies, low-loss high-contrast dielectric structures of the appropriate symmetry can be possibly used for the same purpose. This study has been done for layers providing uniform reflection phases for any point of the surface, and extended also to engineered non-uniform  phase distributions using the physical optics approximation.  It has been shown that a properly designed  single layer of bi-anisotropic particles possessing omega coupling can fully reflect electromagnetic waves coming from different sides with arbitrary phases, and the waves reflected from the two sides of the metamirror can be tailored independently. This concept generalizes the reflectarray concept to the most general phase control of reflections from both sides of the layer using ultimately thin (one layer of dipolar inclusions) structures. In contrast to conventional reflectarrays, metamirrors do not contain a fully reflective ground plane, thus, far from the operational frequency of the mirror these structures are weakly reflective and highly transmitting.

\end{document}